# Influence of Polymer on Shock-Induced Pore Collapse: Hotspot Criticality through Reactive Molecular Dynamics

Jalen Macatangay[1], Chunyu Li[1], and Alejandro Strachan [1]*

[1]School of Materials Engineering and Birck Nanotechnology Center, Purdue University, West Lafayette, Indiana, 47907 USA

* Email: strachan@purdue.edu

## Abstract

The shock initiation of energetic materials is mediated by the localization of mechanical energy into hotspots. These originate through the interaction of the shock and material microstructure; the most potent hotspots are formed by the collapse of porosity. Recent work using molecular dynamics (MD) has shed light on the molecular mechanisms responsible for the shock-to-deflagration transition following pore collapse in pure energetic materials. However, explosive formulations are composites of energetic crystals and a polymer binder, which differs from the prior focus on pure materials. The role of polymer phases in hotspot formation and its criticality remains poorly understood. We use reactive MD simulations to investigate the role of polystyrene and polyvinyl nitrate films around pores in the shock-induced pore collapse of RDX. The polymer affects the hotspots' temperature and their criticality. While the presence of inert polymer often delays or hinders chemical reactions of the energetic material, certain geometries accelerate chemistry. The simulations provide a mechanistic understanding of these phenomena.

# 1. Introduction

The precise characterization of the shock-induced initiation of detonation in high explosives (HEs) is contingent upon a detailed understanding of the formation and reactivity of hotspots. Hotspots, regions of localized energy, accelerate chemical reactions. The onset and rate of these reactions determine whether the hotspot becomes critical and transitions into a deflagration wave.[1] Ultimately, a detonation occurs through the nucleation, growth, and interaction of multiple critical hotspots. During shock loading, hotspots form through energy-localizing interactions between the propagating shockwave and structural defects in the materials microstructure. Several mechanisms are known to result in hotspots. These include pore collapse,[2] adiabatic compression of trapped gases,[3] crack growth,[4] interfacial friction,[5] and localized plastic deformation within shear bands.[6,7] Among these, pore collapse is known to be dominant and control the shock initiation of heterogeneous explosive materials.[8–10]

When a shockwave reaches a pore, the material on the upstream free surface is accelerated and expands into the void. This material becomes recompressed upon collision with the pore's downstream surface, generating localized heating that can prompt chemical reactions.[2] Consequently, accurate measurements of the shock-induced temperature are necessary to quantify the thermal activation of chemistry. Experiments that used optical pyrometry on laser-driven flyer plate shocks have captured the temporal evolution of hotspot temperatures in HEs of varying microstructures.[11–13] However, the ultrafast timescales (ps to ns), spatial localization, and extreme thermodynamic conditions associated with the shock-induced formation of hotspots have hindered a detailed experimental characterization, which is crucial for assessing explosive safety and performance.

For relatively strong shocks, the temperatures of hotspots following shock-induced pore collapse are governed by the pressure-volume (PV) work generated during the downstream impact, which is influenced by the physical pore collapse mechanism.[2] Molecular dynamics (MD) simulations have provided significant insights into these mechanisms to describe the collapse of porosity and the subsequent formation of hotspots. In cylindrical pores, MD studies observed that the pore collapse mechanisms depend on shock strength.[14,15] For weak shocks, pores collapse in a visco-plastic manner, which transitions to hydrodynamic-like behavior for stronger shocks, where the ejected material forms a jet that accelerates into the void. In contrast, shocks on elongated cracks showed more pronounced shock focusing and jetting, even for weak shocks, producing temperatures significantly higher than in cylindrical voids of comparable size.[16] Other factors such as crystallographic orientation, material phase, and pore size can also affect the formation of hotspots.[15,17–19]

For 1D pore geometries (planar gaps resulting from cracks elongated along the shock plane), shock focusing and other complex effects present in 2D/3D defect geometries are removed, leading to expansion/recompression PV work as the primary energy localization mechanism. Holian et al. derived an expression for the maximum expected heating ($\Delta T_{max}$) from a 1D shock-induced pore collapse that scales with the molecular mass ($m$), shock velocity ($U_s$), and particle velocity ($U_p$), $k_B \Delta T_{max} = m U_s U_p / d$, where $k_B$ is Boltzmann's constant and $d$ is the dimensionality.[2] Recent studies revealed that compressive work from the collapse of a 1D pore plays a key role in localizing temperature and potential energy, while adding a shear component can further increase reactivity.[20,21] Reactive force fields (like ReaxFF) enabled the simulation of shock chemistry in HEs. These studies strikingly revealed greater reactivity in hotspots formed from dynamic pore collapse compared to those of equivalent size and temperature but created under equilibrium conditions.[22] This was later attributed to the localization of latent potential energy in plastically deformed molecular states that accelerate chemical reactions.[17,23,24]

Most of the prior MD research has centered on the shock response of pure energetic material components. Yet, explosive materials primarily exist as polymer-bonded explosives (PBXs), energetic composites consisting of explosive crystals embedded within a polymer binder. The binder significantly influences explosive sensitivity; chemically inert polymers soften external impacts, while energetic polymers can enhance the overall energy release.[25,26] Under high-velocity impacts, polymers can exhibit complex response mechanisms such as virtual melting and dynamic glass transitions, which can complicate the shock to detonation process in PBXs.[27,28] The microstructures of PBXs, which feature voids, cracks, and other defects between the binder and HE components, also influence the localization of thermal and strain energy.[17,29,30]

Recent laser-driven flyer plate experiments on nominally defect-free HMX/PU composites observed the generation of hotspots at the binder-HE interfaces, particularly where the polymer is misaligned with the edges and corners of HMX.[31,32] Numerical simulations attributed this to enhanced energy localization from shock focusing occurring at the sharp corners. Furthermore, these calculations identified that the collapse of voids within the binder can induce reactions in the adjacent HMX for strong shocks.[33] Nonreactive MD pore collapse simulations on TATB with polyethylene coated on both sides of a planar gap found that the material impedance differences resulted in different shock states that lowered core hotspot temperatures.[26] It is evident that a detailed understanding of hotspot formation mechanisms and their evolution under shock loading in PBXs is still not well-grounded. Specifically, the effect of the presence and arrangement of polymers surrounding voids on the shock-to-deflagration transition following the collapse of porosity is still unclear.

In response, we utilize a 1D pore collapse geometry with polymer films located around the gap surfaces. For these simulations, the formation and criticality of hotspots are only contingent on the shock speed and pore surface material. Therefore, we can elucidate the effect of polymers surrounding voids on the localization of energy and the transition to deflagration. Using reactive MD, we also assess how exothermic chemistry is influenced by the pore surface structure. These simulations show that while the presence of polymer impacts the hotspots' temperatures, certain geometries accelerate chemistry, leading to a faster transition to deflagration. The generality of these findings is evaluated through simulation of two different polymer binders that differ in chemical nature.

## 2. Methods

We perform reactive, all-atom MD simulations using the LAMMPS software package[34] and the ReaxFF-2018 force field.[35] ReaxFF is an empirical force field that dynamically adjusts atomic interactions based on partial bond orders that are dependent on local atomic environments.[36] Numerous studies have confirmed the accuracy of ReaxFF to describe shock-induced chemistry, reaction kinetics, and thermal decomposition pathways in energetic materials.[37–40] Most recently, the ReaxFF-2018 force field adds a low-gradient (lg) attractive term[41] to ReaxFF-2014[42] that corrects long-range London dispersions and improves bulk property predictions of energetic materials. This has predicted the Hugoniot shock states, denotation velocities, and CJ pressures in liquid nitromethane and TATB,[43,44] as well as shock initiation chemistry in various energetic materials.[45]

Our systems closely resemble the models used in Ref. [20], which contained a planar (1D) gap centered between two crystalline RDX slabs. All simulations were conducted with a timestep of 0.1 fs, with environment-dependent partial atomic charges calculated at every timestep using a charge equilibration tolerance of $10^{-6}$. To create the pure RDX (no polymer) system, we first relax a 3x3x3 α-RDX supercell using the conjugate gradient minimization scheme, followed by equilibration under isothermal-isobaric (NPT) conditions at 300 K and 1 atm for 50 ps. The resulting density and lattice parameters are as follows: 1.64 g/cm$^3$, a = 13.92 Å, b = 11.22 Å, c = 11.43 Å. Next, we replicate the system to a length of ~360 nm in the shock propagation direction ([001]). Nonperiodic boundaries were implemented along this direction, while the lateral boundaries were kept periodic. A 40 nm gap was produced by removing unit cell proportions of molecules beginning at the center of the system, creating two RDX crystals, each with an initial length of ~160 nm. The system was further equilibrated under isothermal-isobaric (NVT) conditions at 300 K for 25 ps, resulting in the model displayed in Figure 1.

Polystyrene (PS) and polyvinyl nitrate (PVN) films were constructed using the Polymer Modeler nanoHUB tool.[46] An initial bulk system of 40 monomer chains was generated with a density of 0.5 g/cm$^3$. Using the Dreiding force field,[47] this was first energy-minimized and equilibrated under NVT conditions for 50 ps at 600 K, followed by NPT conditions for 200 ps at 600 K and 1 atm. The relaxed bulk system was then cooled to 300 K using a cooling rate of 20 K/200 ps. To quantify the mismatch in mechanical properties between RDX and PS, we provide Hugoniot relations in the shock velocity vs particle velocity ($U_s$-$U_p$), pressure vs volume (P-V), and temperature vs particle velocity (T-$U_p$) planes in Section SM-1 of the Supplemental Material. These shocks were conducted uniaxially along [001] on the bulk systems using the Hugoniotstat method. Additionally, we include tensile and compressive stress-strain responses at 300 K and a strain rate of $2\times10^9$ s$^{-1}$ in Section SM-2 of the Supplemental Material.

We then create a slab by introducing free surfaces in the shock direction and deforming the lateral dimensions under NVT/sllod conditions to align with those of the RDX crystal. Final relaxation is performed at 300 K under NVT conditions for 200 ps, followed by an additional 100 ps under ReaxFF. The resulting densities for the PS and PVN slabs (thicknesses of 3, 5, and 10 nm) were respectively around 1.00 and 1.55 g/cm$^3$, which align well with experiments.[48–50] A polymer slab was then positioned at the upstream or downstream surface of the gap after translating one of the RDX crystals outwards to maintain the 40 nm gap width. We additionally created an RDX-PS system with ~10 nm PS slabs on both surfaces to further evaluate the interplay between upstream and downstream polymer collapse mechanisms. All final systems were equilibrated at 300 K for 25 ps under NVT conditions to relax the newly created RDX-polymer interfaces. Figure 1 illustrates the variations in the pore surface structure for these simulations compared to the pure RDX system.

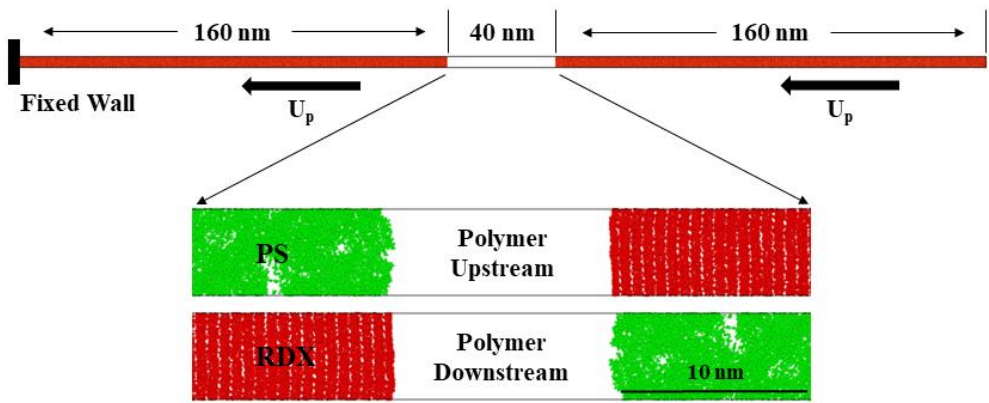

*Figure 1: Simulation setups for a 1D shock-induced pore collapse. The full schematic represents the no polymer (pure RDX) system, while the close-up of the void highlights the configurations with 10 nm of polymer along the upstream and downstream sides of the pore.*

Shock loading was conducted using the reverse ballistic technique,[51] which assigns a particle velocity on all atoms towards a rigid wall under adiabatic (NVE) conditions. In these models, relief waves generated from the initial shockwaves reaching the cell boundaries limit the time scales for analyzing hotspots. Thus, we apply Shock Trapping Internal Boundaries (STIBs) at a time after pore collapse when the initial shockwaves are well beyond the hotspot region. STIBs involve fixing atom layers to act as boundaries that isolate the hotspot, allowing for extended hotspot evolution.[23] Local temperatures are then determined through an Eulerian and Lagrangian binning analysis. As described in Ref. [22], we compute bin temperatures after taking the difference between the atomic velocities and translational center of mass velocities of each bin.

## 3. Role of an inert polymer on shock-induced pore collapse reactivity

### 3.1 RDX baseline

We characterize the role of polymer film on hotspot criticality through the evolution of temperature and density within the hotspot region. Hotspots become critical when the heat generation from product forming chemical reactions overcomes the dissipation due to thermal diffusion, rapidly increasing local temperatures. Prior reactive pore collapse simulations on 40 nm cylindrical voids in RDX at $U_p$ = 2.0 km/s observed substantial temperature rises from ~2,000 K up to 4,000 K due to the formation of final product species.[22] Similar temperatures were reported for the collapse of a 40 nm planar gap in RDX, though its critical particle velocity is narrowly higher ($U_p$ = 2.1 km/s).[20] We reference these findings as a basis to identify critical hotspots in terms of temperature.

Figure 2 displays position-time (x-t) diagrams along the shock propagation direction and relative to the time of pore collapse ($t_0$) for the pure RDX system at $U_p$ = 1.8 and 1.9 km/s. The first and second RDX crystal slabs are designated as (I) and (II), respectively, with the void separating them denoted as (a). For times before the collapse of porosity ($t < t_0$), the initial shockwave travels through and shocks the first crystal (b). When the shockwave reaches the upstream free surface, a reflective wave is formed as material expands freely into the gap. At $t_0$, the expanded material impacts the initially unshocked second crystal; the cross-section where this collision occurs is defined as the impact plane (c). This generates a subsequent reflection that recompresses the expanded material and elevates local temperatures. Additionally, a forward-directed shockwave propagates into the second crystal.

The temperatures and pressures in the hotspot region immediately after pore collapse ($t_0 + 1$ ps) are approximately 1050 K and 15 GPa for $U_p = 1.8$ km/s and 1160 K and 18 GPa for $U_p = 1.9$ km/s. These marginally increase over time as the shockwaves propagate away from the impact plane. At around $t_0 + 50$ ps, the hotspot in the case of $U_p = 1.9$ km/s significantly heats, signifying the onset of self-sustaining chemical reactions that produce a deflagration wave (d). In contrast, the hotspot quenches when $U_p = 1.8$ km/s. The $U_p = 1.9$ km/s critical particle velocity required for deflagration is slightly lower than the value predicted with ReaxFF-2014 ($U_p = 2.1$ km/s).[20] This is due to the more accurate (higher) density and exothermicity of ReaxFF-2018. Furthermore, the higher sensitivity of ReaxFF-2018 relative to 2014 is observed in the shock initiation of nitromethane.[43] Therefore, we use $U_p = 1.9$ km/s, near the threshold of criticality, to assess the effect of polymer films in the simulations discussed in the following sections.

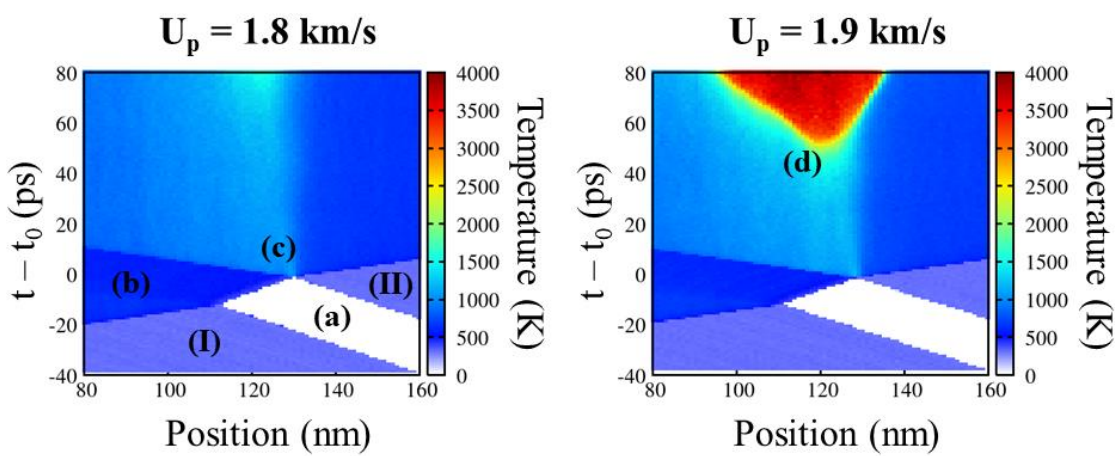

*Figure 2: x-t diagrams of temperature for pure RDX for particle velocities ($U_p$) of 1.8 km/s and 1.9 km/s relative to the time of pore collapse ($t_0$).*

**3.2 Role of polystyrene films**

The inclusion of polymer films along the pore surfaces gives rise to markedly different hotspot behaviors following the collapse of porosity. Figure 3 shows temperature and density x-t diagrams for two RDX-PS systems, where a 10 nm PS film is located on either the downstream or upstream surface (see Figure 1 for simulation setups). Density profiles are included to highlight the differences between the polymer and RDX, as well as the progression of chemistry over time. These are also compared to the pure RDX (no PS) system at $U_p = 1.9$ km/s, as shown in Figure 2. Additional x-t diagrams for simulations with 3 and 5 nm PS films are featured in Section SM-3 of the Supplemental Material. Starting with the PS downstream case, the pore collapse process is characterized by the expansion and recompression of RDX into the unshocked PS film along the second RDX crystal. The lower density PS (~1.0 g/cm$^3$) attenuates the impact, leading to peak pore collapse temperatures approximately 100 K cooler relative to the no PS case (see PS downstream temperature diagram). Furthermore, the higher compressibility of PS under shock, as seen in Figure S1b, dissipates some of the generated PV work. Over time, the temperature profiles remain steady, even well after pore collapse, and we do not observe the development of a deflagration wave. Comparison of its density plot to that of the no PS case shows that the PS film inhibits the formation of reaction products that cause local volumetric expansion.

In the PS upstream case, the polymer expands into the void and collides with the downstream RDX section. During expansion, the more compliant PS reaches a significantly lower density than the expanding RDX in the pure RDX, baseline case before recompression, as seen in its density x-t diagram. As will be quantified below, this leads to larger PV work during recompression as compared to the other cases, giving rise to higher temperatures in the PS region following pore collapse (~1570 K at $t_0$ + 1 ps). Shockwave interactions with the higher impedance RDX at both the upstream and downstream interfaces generate multiple sets of reflection and transmission waves. These result in local pressure elevations comparable to the no PS case. Remarkably, the RDX region directly upstream from the PS film reaches the peak temperatures of the polymer, and approximately 35 ps after impact, we observe reactions and the development of a deflagration wave. This becomes evident from the temperatures notably exceeding those of the polymer and the densities decreasing to around 2.0 g/cm$^3$ within the hotspot region, signifying the formation of product gases. Both x-t diagrams show that the critical hotspot originates in the RDX material directly behind the PS film and generates a deflagration wave propagating opposite to the initial shockwave. We also observe that the PS film obstructs deflagration into the downstream RDX crystal. The inert polymer does not significantly heat or react within our simulation timescales. Simulations of PS films of various thickness (Section SM-3 in the Supplemental Material) show the acceleration of reactions with thicker PS films on the upstream surface, whereas a thicker PS film on the downstream surface retards it.

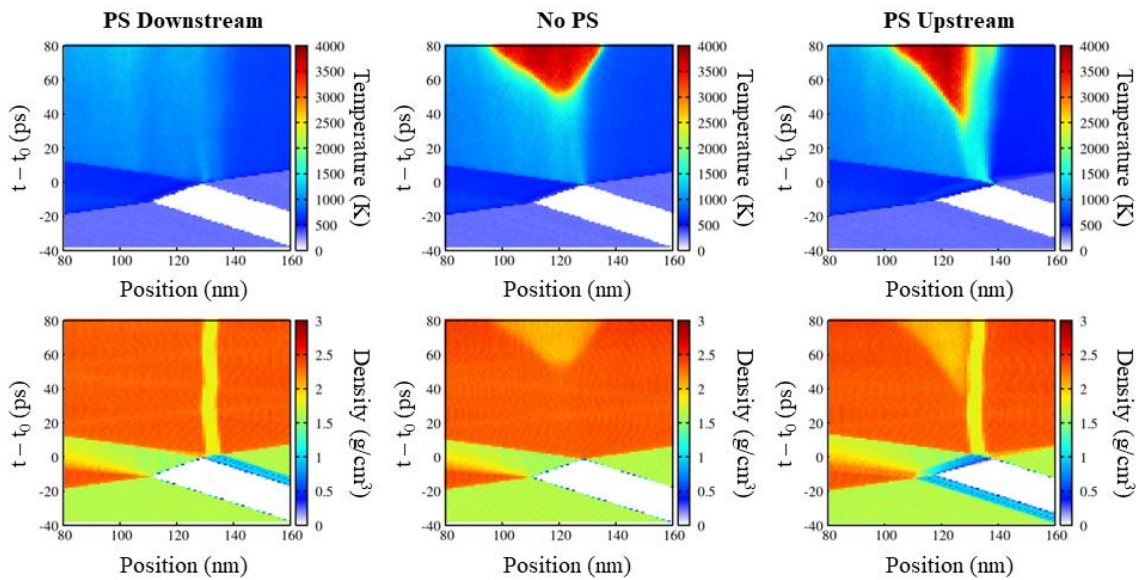

*Figure 3: x-t diagrams of temperature (top) and density (bottom) for the no PS, PS downstream, and PS upstream cases.*

The profound differences in the temperature rise after pore collapse between the PS upstream and no PS cases are primarily due to the PV work done during the recompression of the expanding material. To quantify the disparities during the expansion and recompression of upstream RDX and PS, we first show the density profiles along the shock direction at $t_0$ in Figure 4a. This is equivalent to a horizontal line at $t_0$ across the density x-t diagrams in Figure 3. We note that the expansion of the PS downstream case is identical to the no PS case and is therefore not shown.

The shocked RDX near the fixed wall (position = 0 nm) is at a density of ~2.4 g/cm$^3$, which decreases to around its initial density (1.64 g/cm$^3$) due to the propagation of the rarefaction fan into the shocked material as the material expands into the void. The leading edge of the expanded material is expected to be faster than the average rarified material, which is seen with its lower density of ~1.5 g/cm$^3$ for the no PS case (red curve). In contrast, the softer polymer in the PS upstream case (green curve) expands to densities as low as ~0.7 g/cm$^3$.

We now estimate the local heating based on the PV work done due to recompression following the work of Holian et al.[2] Figures 4b and 4c display the time evolution of upstream material adjacent to the gap in pressure-volume space. These describe the evolution of RDX (panel b) and PS (panel c) during the pore collapse process including the initial shock (1), expansion (2), and recompression (reshock) against the downstream surface (3). The PV work of recompression corresponds to the area between curves (2) and (3), the release isentrope and reshock Rayleigh line, which is then divided by the specific heat to obtain the increase in temperature. We used the classical limit for the specific heat of RDX and PS, which are 2358 and 3832 J·kg$^{-1}$·K$^{-1}$, respectively. The estimated heating from the PV work due to RDX expansion and recompression is around 850 K, notably lower than the increase in temperature from the recompression of PS (~1360 K) and in good agreement with our simulations. The difference is a direct result of the lower stiffness of PS, which results in more expansion prior to recompression. We additionally compare these values to the theoretical maximum temperature increase ($\Delta T_{max}$) derived by Holian et al.,[2] assuming full expansion. For RDX and PS at $U_p$ = 1.9 km/s, $\Delta T_{max}$ equals around 4980 and 2790 K, respectively. In the case of the stiffer RDX, the observed temperatures for our 40 nm planar gap are between 20% and 25% of its $\Delta T_{max}$. In comparison, the collapse of circular voids at similar shock strengths can achieve 30% to 40% of their $\Delta T_{max}$ due to shock focusing effects that result in more rarefication of the expanding gas.[16] For elongated cracks in the shock direction, temperatures can reach significantly higher values that are close to its $\Delta T_{max}$. Interestingly, the expanding PS reaches up to 65% of its $\Delta T_{max}$ without focusing effects, which indicates its enhanced capability to accelerate and rarify in void space to generate more PV work.

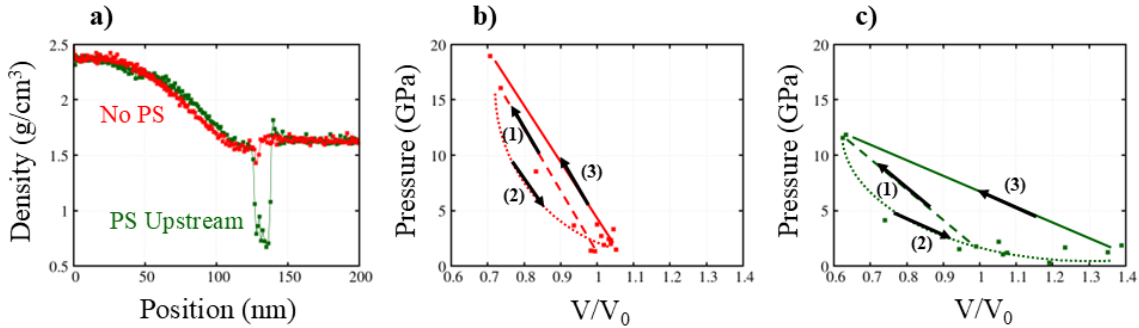

Figure 4: a) Density profiles at the time of pore collapse ($t_0$) and pressure-volume curves of b) the upstream RDX in the no PS case and c) the polymer in the PS upstream case. The numbers and arrows denote the time evolution in pressure-volume space as follows: 1) initial shock Rayleigh line, 2) expansion into the void, and 3) reshock Rayleigh line due to impact against the downstream surface.

To further investigate the phenomena influencing the shock-to-deflagration transition observed in the PS cases, we implement a Lagrangian binning framework to track the temperatures of groups of atoms near the impact plane. Figure 5 shows the temperature evolution of these groups color-coded by their proximity to the impact plane, with darker curves representing bins that are closer.

Additionally, Section SM-4 of the Supplemental Material displays simulation snapshots of the regions described in Figure 5, colored by the Lagrangian binned temperature. For the no PS and PS downstream cases, each curve corresponds to a bin of 1,512 RDX atoms in an ~22 nm region (~16 nm when recompressed) of the upstream material. We observe that the time evolution of temperatures highlights the trends discussed in Figure 3. Beginning with the no PS case, the first shock results in a temperature of ~650 K. This temperature is lower for material closer to the free surface due to the immediate expansion following the initial shock. Recompression leads to a hotspot temperature of approximately 1160 K and reactions nucleate roughly 10 nm away from the impact plane (intermediate color-shaded bins). With the downstream case, the temperature profiles of the initial shock and expansion are essentially identical. During recompression, the lower-density PS layer in the downstream section compresses more readily than the RDX (see Figure S1b), resulting in only a gradual temperature increase of the RDX bins. The different reshock state of the collapsing RDX produces a hotspot temperature ~100 K cooler than the no PS case and does not heat rapidly enough to reach a critical state.

In the PS upstream case, the expanding material consists of both polymer and RDX. The five bins with the darkest shading denote the PS film, each containing 1,284 atoms and a recompressed bin width of ~0.7 nm. Unlike the previous cases, the shock due to recompression leads to diverse temperature increases, with more pronounced heating in the bins closer to the impact plane, aligning with the PS region. Quite interestingly, while the initial shock temperature of these PS bins is higher, the temperatures gradually decrease over a 30 ps timeframe following pore collapse. This aligns in time with the continuous heating of the lighter-shaded RDX bins until the rapid rise in temperature signifies a critical hotspot. For the PS region near the upstream RDX interface (intermediate color-shaded bins), we observe steady heating that does not approach a critical temperature. Thus, these findings indicate that the presence of an inert polymer along the upstream surface of the pore expedites deflagration by transporting heat to the recompressed RDX. This additional source of heating contributes to the earlier initiation of exothermic chemistry observed in our simulations.

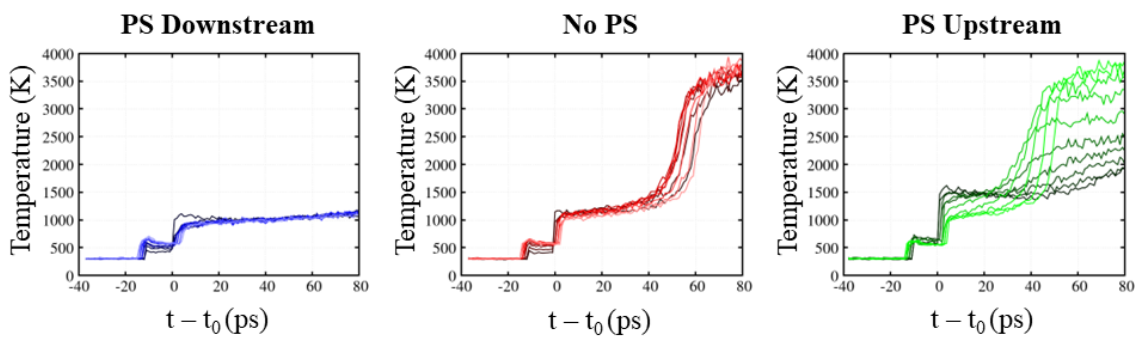

*Figure 5: Temperature-time profiles of the upstream material for the no PS, PS downstream, and PS upstream cases. Each curve represents a Lagrangian bin color-shaded according to its proximity to the impact plane, with darker shades indicating closer proximity. In the PS upstream case, the five darkest-shaded curves correspond to PS material.*

Lastly, we consider the scenario with PS films on both surfaces of the gap to compare the relative potency of the upstream and downstream polymer collapse mechanisms. This setup resembles intergranular pores located within the binder and adjacent to HE crystals, which have also been identified as potential sites for hotspot formation.[33] Figure 6 displays temperature and density x-t diagrams of the RDX-PS shock simulation with 10 nm of PS along each surface. The collision between PS films generates peak temperatures of ~1300 K concentrated in the polymer. This is around 300 K lower than the upstream case, yet approximately 200 K higher than the no PS case. The additional heating due to the increased PV work done by the expanding polymer is partially lessened by the downstream polymer. Furthermore, the compression of twice as much polymer material (20 nm) relative to the upstream case reduces the strength of the reshock wave into the collapsing RDX. The lower shock temperatures in the polymer reduce the rate of heat transport, delaying the transition to criticality. Eventually, we observe a critical hotspot in the recompressed RDX around 75 ps after the collapse of porosity. Our findings indicate that the energy localization from the upstream polymer is greater than the retardation from the downstream polymer. In this case, the localized energy is adequate to cause ignition in the adjacent HE material, though this pore configuration is less sensitive compared to the pores in the upstream and no PS cases.

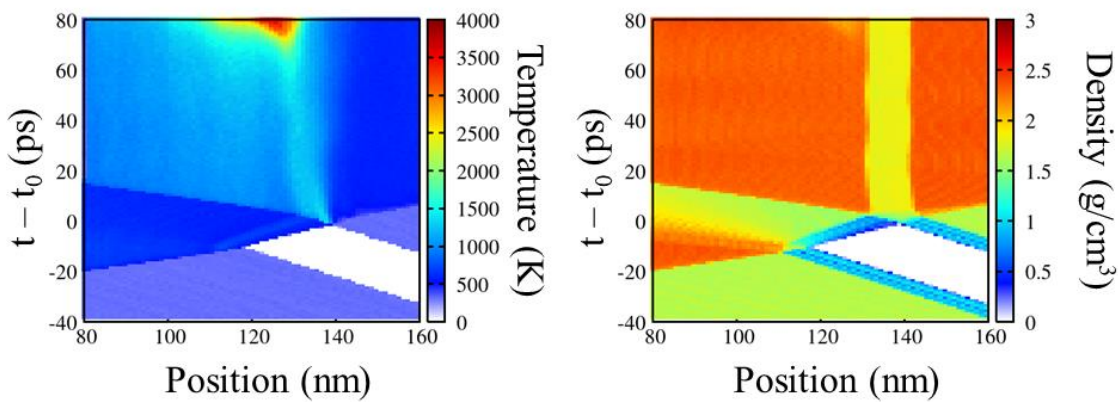

*Figure 6: x-t diagrams of temperature and density for 10 nm of PS on both surfaces of the pore.*

## 4. Role of a reactive polymer on shock-induced pore collapse reactivity

Our studies have revealed that polystyrene along the surface of the pore significantly varies the onset of hotspot criticality and chemical reactions. Yet, it remains uncertain whether similar effects are seen for different polymer binder formulations. To extend our analysis, we consider PVN, a high energy density polymer of major interest in explosive and propellant applications. In contrast to PS, PVN exhibits increased chemical reactivity due to the presence of nitrogen and oxygen as additional oxidizers, which can intensify the overall energy release.[52] Prior reactive MD simulations identified the formation of $NO_2$ as the primary reaction path in PVN, yielding final product species resembling those found in RDX and other HE compounds.[49] Moreover, we expect a lower impedance mismatch effect due to similar densities between PVN (1.55 g/cm$^3$) and RDX (1.64 g/cm$^3$).

Figure 7 presents the temperature and density x-t diagrams for the scenarios previously discussed with PS, but now with PVN as the polymer film. Additionally, Section SM-5 in the Supplemental Material includes the diagrams for 3 and 5 nm PVN simulations. In the downstream case, the RDX expansion into the PVN surface leads to an initial hotspot temperature of ~1080 K, which is slightly cooler, yet comparable to the no PVN case. Unlike the PS downstream case, the PVN film undergoes sustained heating and eventually reacts, forming a critical hotspot originating at the upstream RDX – downstream PVN interface. The temperature evolution of the hotspot closely follows that of the no PVN case, as both exhibit a critical hotspot forming around $t_0 + 50$ ps with deflagration waves propagating in both directions. For the PVN upstream case, the presence of PVN results in an impact that generates a temperature of ~1200 K and a critical hotspot within 20 ps. This accelerated deflagration relative to other pore surface configurations is consistent with the PS upstream case. However, the most notable distinction is that PVN significantly expedites the production of exothermic products, rapidly increasing temperatures in the first few picoseconds within the reshocked PVN region. We observe that the reactive properties of PVN enhance the system's overall exothermicity following pore collapse, leading to critical hotspots in all cases.

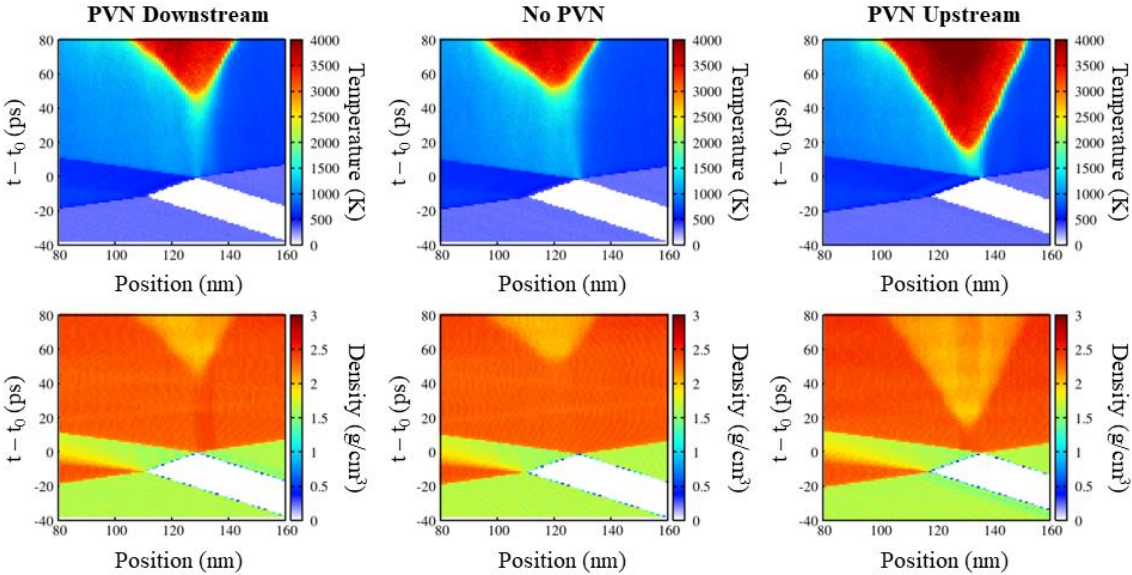

*Figure 7: x-t diagrams of temperature and density for the no PVN, PVN downstream, and PVN upstream cases.*

## 5. Conclusions

We used reactive molecular dynamics to investigate how the presence of polymer-coated pore surfaces influences hotspot formation and criticality in a shock-induced pore collapse simulation. We implement a 1D simulation setup with two RDX crystals separated by a 40 nm gap, specifically designed to isolate the effects of pore surface material on energy localization. Our simulations reveal markedly different shock-to-deflagration behaviors based on the polymer type and geometry. We find that polystyrene, a chemically inert polymer, on the upstream surface of the pore accelerates the formation of a critical hotspot. Its greater compressibility relative to RDX results in larger PV work during recompression and thermal transport from the polymer to the collapsing RDX further facilitates rapid heating. The deflagration front propagates into the reshocked RDX, with the polymer preventing reactions from taking place in the downstream crystal. In contrast, PS along the downstream surface of the pore mitigates the impact from pore collapse, diminishing the strength of the compression wave into the RDX and retarding deflagration. For cases with polyvinyl nitrate, reactions within the polymer contribute to critical hotspots in all cases. Product-forming chemistry occurs in the first few picoseconds following the collapse of porosity with PVN on the upstream surface, significantly elevating hotspot temperatures. Overall, these trends emphasize the importance of polymers in explosive formulations on hotspot criticality, a factor that has been neglected in atomistic modeling. Our results encourage further exploration of more complex microstructures to improve simulations of hotspot dynamics for better informed continuum models.

## Supporting Information

RDX/PS shock Hugoniots, isothermal stress-strain responses, and Lagrangian binned temperature simulation snapshots, as well as additional x-t diagrams for intermediate polymer thicknesses

## Acknowledgments

The research was sponsored by the Army Research Office and was accomplished under Cooperative Agreement Number W911NF-22-2-0170. The views and conclusions contained in this document are those of the authors and should not be interpreted as representing the official policies, either expressed or implied, of the Army Research Office or the U.S. Government. The U.S. Government is authorized to reproduce and distribute reprints for Government purposes notwithstanding any copyright notation herein. We acknowledge computational resources from nanoHUB and Purdue University through the Network for Computational Nanotechnology.

# Supplemental Materials to:
# Influence of Polymer on Shock-Induced Pore Collapse: Hotspot Criticality through Reactive Molecular Dynamics

Jalen Macatangay[1], Chunyu Li[1], and Alejandro Strachan [1]*

[1]School of Materials Engineering and Birck Nanotechnology Center, Purdue University, West Lafayette, Indiana, 47907 USA

* Email: strachan@purdue.edu

## SM-1. RDX and PS Shock Hugoniots

Figure S1 displays the RDX and PS Hugoniot relations for uniaxial shocks in the [001] direction. The shocks were performed on the respective bulk system initially at 300 K and 1 atm using the Hugoniotstat method for pressures between 2 and 25 GPa. The initial densities ($\rho_0$) of the RDX and PS systems are 1.66 and 1.15 g/cm$^3$. All variables of interest were averaged over the last 10 ps of a 25 ps simulation.

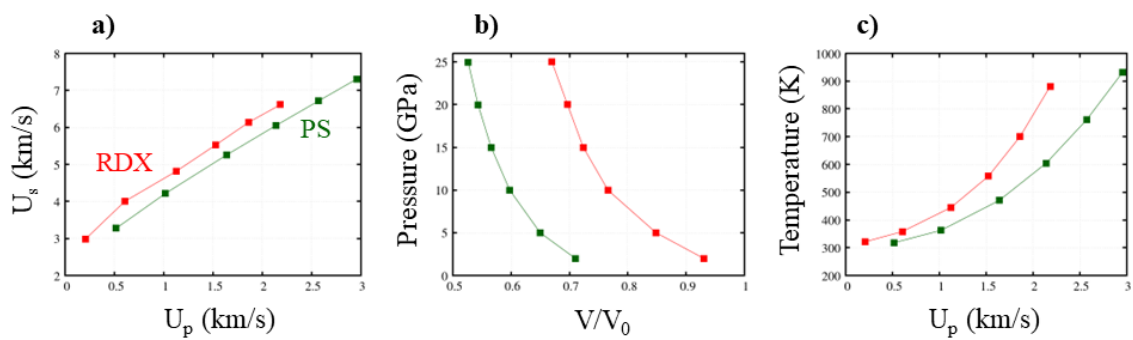

*Figure S1: a) $U_s$-$U_p$ b) P-V and c) T-$U_p$ Hugoniot equations of state, calculated from a uniaxial [001] shock with the Hugoniotstat method.*

## SM-2. RDX and PS Mechanical Responses

Figure S2 displays the uniaxial tensile and compressive stress-strain responses for RDX and PS in the [001] direction. The bulk systems were deformed at a constant temperature of 300 K and a strain rate of $2\times10^9$ s$^{-1}$.

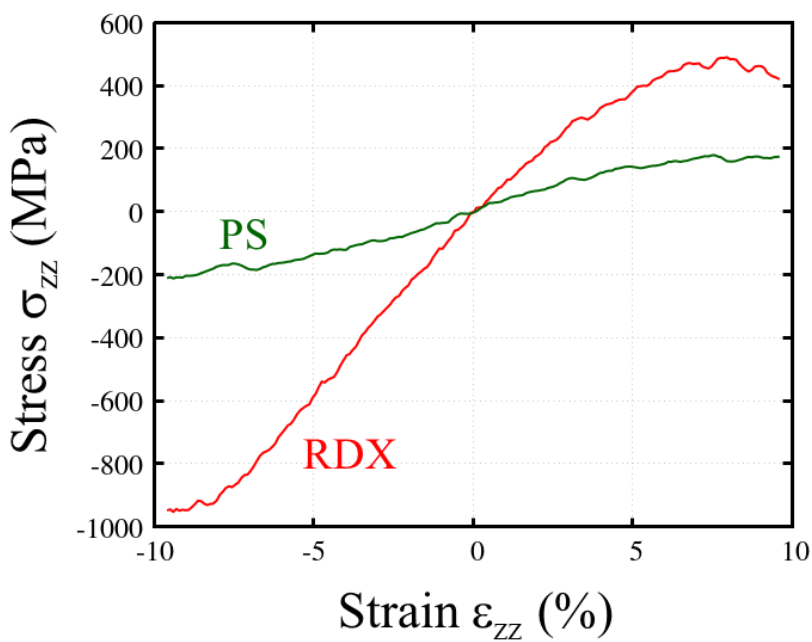

*Figure S2: Stress-strain curves under compression and tension for RDX and PS.*

## SM-3. Temperature Profiles of Intermediate PS Thicknesses

Figure S3 shows temperature x-t diagrams for pore collapse simulations with 3 and 5 nm PS films on the upstream and downstream surfaces. This figure supplements the results seen in Figure 3 of the main manuscript to highlight the trends of increasing PS thickness.

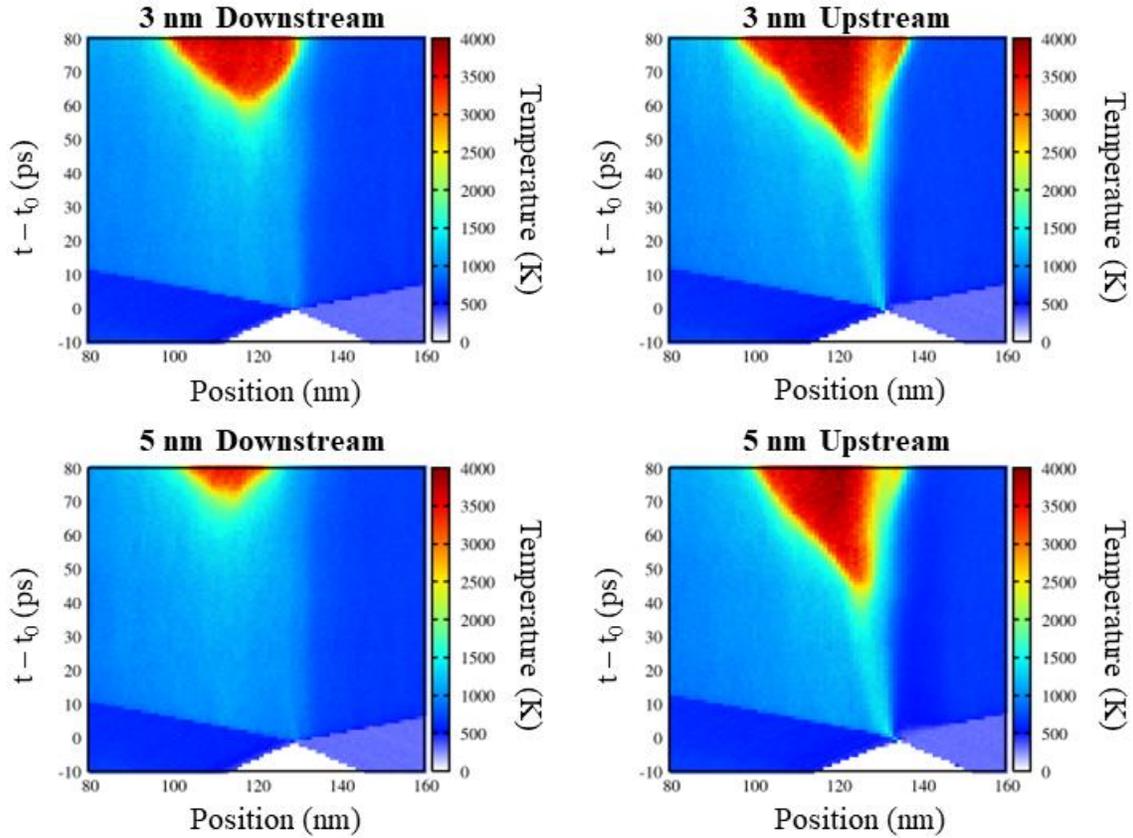

*Figure S3: Temperature x-t diagrams for 3 nm and 5 nm PS cases on the downstream and upstream surfaces of the pore.*

## SM-4. Lagrangian Binned Temperature Simulation Snapshots

Figure S4 displays a ~20 nm region of the upstream material directly behind the impact plane. The atoms are colored by their Lagrangian binned temperature and shown for times following the collapse of porosity ($t_0$).

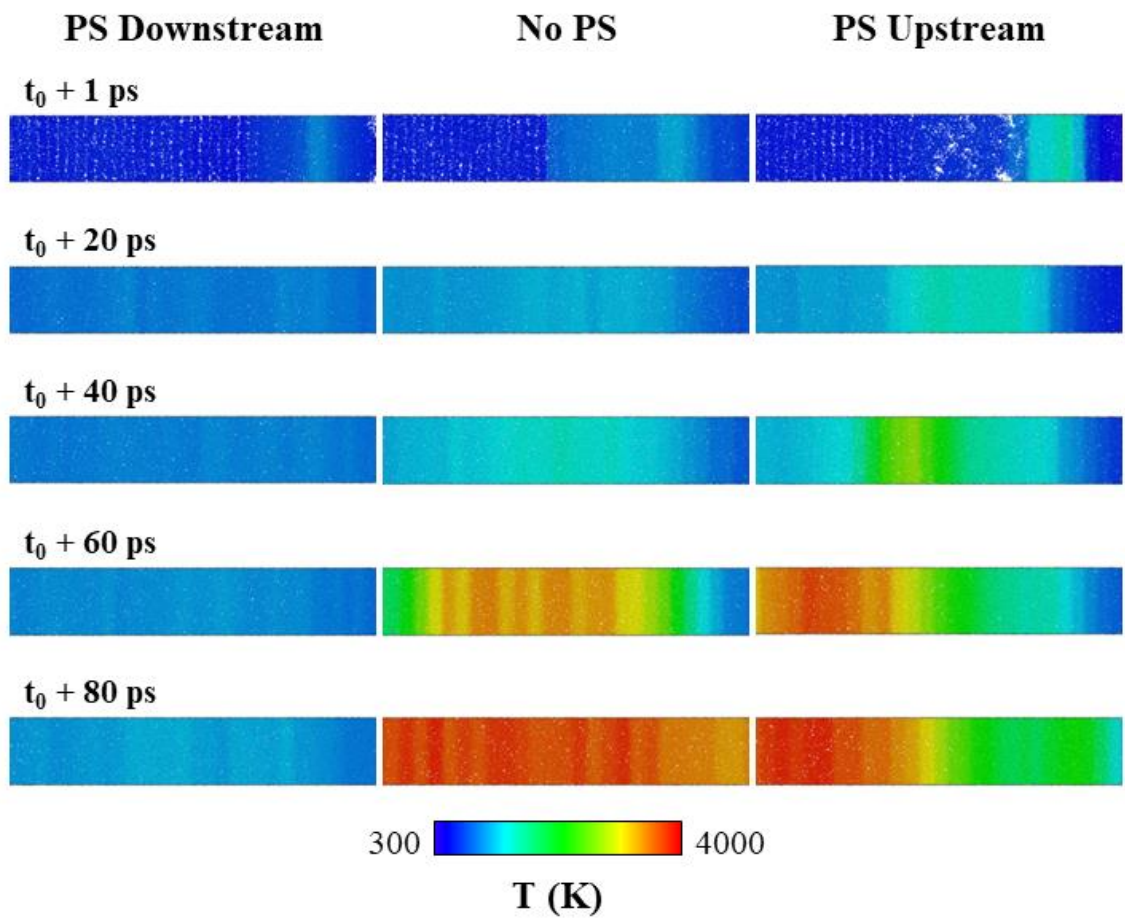

*Figure S4: Simulation snapshots of the PS downstream, no PS, and PS upstream cases colored by temperature.*

## SM-5. Temperature Profiles of Intermediate PVN Thicknesses

Figure S5 shows temperature x-t diagrams for pore collapse simulations with 3 and 5 nm PVN films on the upstream and downstream surfaces. This figure supplements the results seen in Figure 7 of the main manuscript to highlight the trends of increasing PVN thickness.

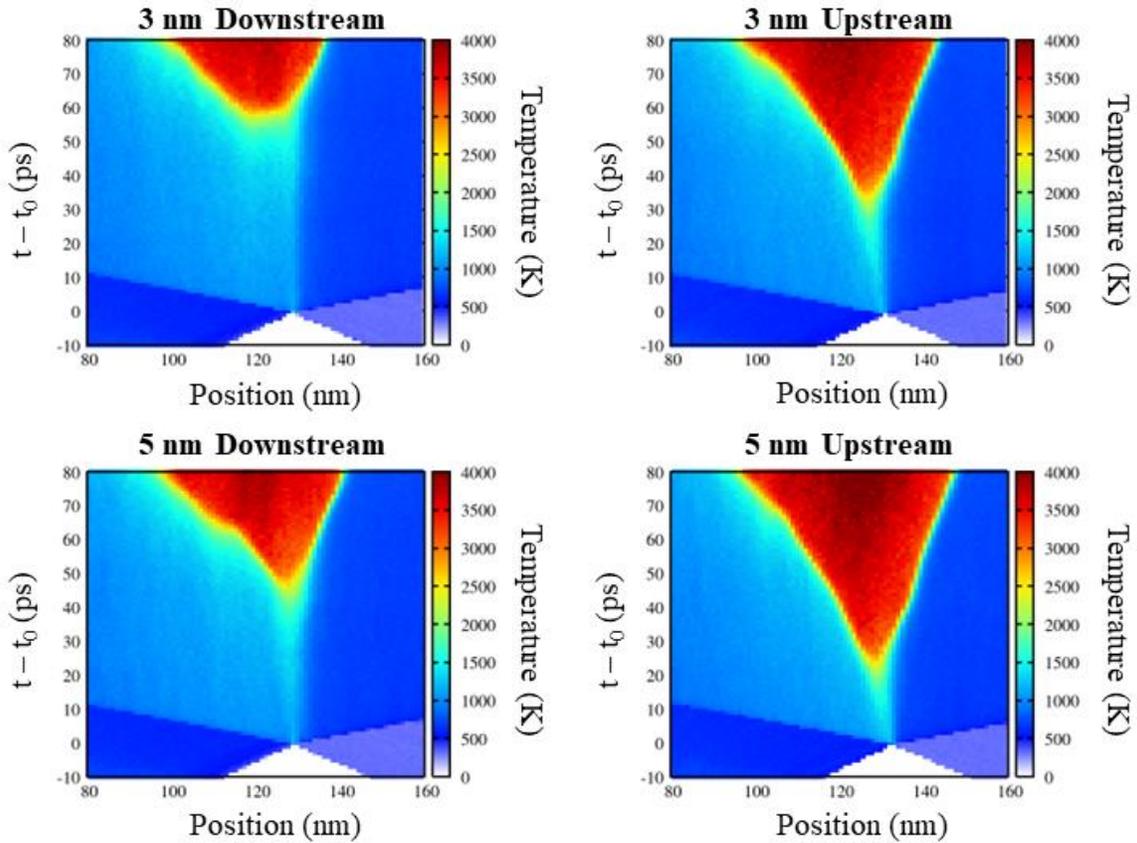

*Figure S5: Temperature x-t diagrams for 3 nm and 5 nm PVN cases on the downstream and upstream surfaces of the pore.*